\documentclass[aps,prb,reprint,twocolumn,groupedaddress,showpacs]{revtex4-1}

\usepackage{graphicx}
\usepackage{amsmath}
\usepackage{hyperref}
\usepackage{soul}
\usepackage{color}

\begin{document}

\title{Anomalous Josephson effect in semiconducting nanowires as a signature of the topologically nontrivial phase}

\author{Konstantin N. Nesterov}
\author{Manuel Houzet}
\author{Julia S. Meyer}
\affiliation{Univ. Grenoble Alpes, INAC-PHELIQS, F-38000, Grenoble, France}
\affiliation{CEA, INAC-PHELIQS, F-38000, Grenoble, France}

\date{\today}

\begin{abstract}
We study Josephson  junctions made of semiconducting nanowires with Rashba spin-orbit coupling, where superconducting correlations are induced by the proximity effect. In the presence of a suitably directed magnetic field, the system displays the anomalous Josephson effect: a nonzero supercurrent in the absence of a phase bias between two superconductors. We show that this anomalous current can be increased significantly  by tuning the nanowire into the helical regime. In particular, in a short junction, a large anomalous current is a signature for topologically nontrivial superconductivity in the nanowire.
\end{abstract}


\maketitle

\setcounter{topnumber}{1} 

\section{Introduction}\label{Sec_intro}

Originally discussed in  particle physics, Majorana fermions have recently attracted a  lot  of attention from the condensed-matter community.~\cite{Wilczek2009,Alicea2012,Elliott2015} Several experiments report possible signatures of Majorana bound states,~\cite{Mourik2012,Das2012,Deng2012,Churchill2013,Finck2013,Rokhinson2012,Nadj-Perge2014,Xu2014} although additional evidence is still needed.  
One of the most popular solid-state systems to search for such states is a semiconducting nanowire with strong Rashba spin-orbit coupling (SOC). When a superconducting gap  is induced by the proximity effect, a Zeeman field applied along the nanowire can drive the system into a topologically nontrivial phase, in which  Majorana bound states are predicted  at its ends.~\cite{Lutchyn2010,Oreg2010}
In this phase, the system is effectively spinless, namely spin is correlated with the direction of motion.

Here we propose that the anomalous Josephson effect (AJE) may be used to probe whether the nanowire is in the topologically nontrivial phase. In this effect, a finite supercurrent $I_{\rm an}$ flows between two superconductors at zero phase difference. Such an anomalous current has been predicted in a variety of topologically trivial systems, mainly due to an interplay between spin-orbit coupling and a Zeeman field applied
to the junction area,~\cite{Krive2004,Reynoso2008,Buzdin2008,Zazunov2009,Yokoyama2014-PRB,Mironov2015} but also in hybrid structures with noncoplanar ferromagnets~\cite{Braude2007} (see also Ref.~\onlinecite{Eschrig2015} for a recent review).
 In the former case, the anomalous current results from a magnetoelectric coupling~\cite{Edelstein1996} between the electric field responsible for the Rashba SOC and the Zeeman field. The Rashba SOC separates the spectrum into two helical bands, and the effect is typically small because of the competition between them. A much larger effect can be obtained in a topological system, where only one band exists at the Fermi level.~\cite{Tanaka2009,Linder2010, Black-Schaffer2011,Ojanen2013, Nussbaum2014,Lu2015,Marra2015,Zyuzin2015}
 Recently, it was demonstrated that the maximal effect in a short junction based on the helical edges states of a quantum spin-Hall insulator (QSHI) is found when the Zeeman field is applied not only to the junction area, but to the superconductors as well.~\cite{Dolcini2015}  
 
\begin{figure}[t]
 \includegraphics[width=240pt]{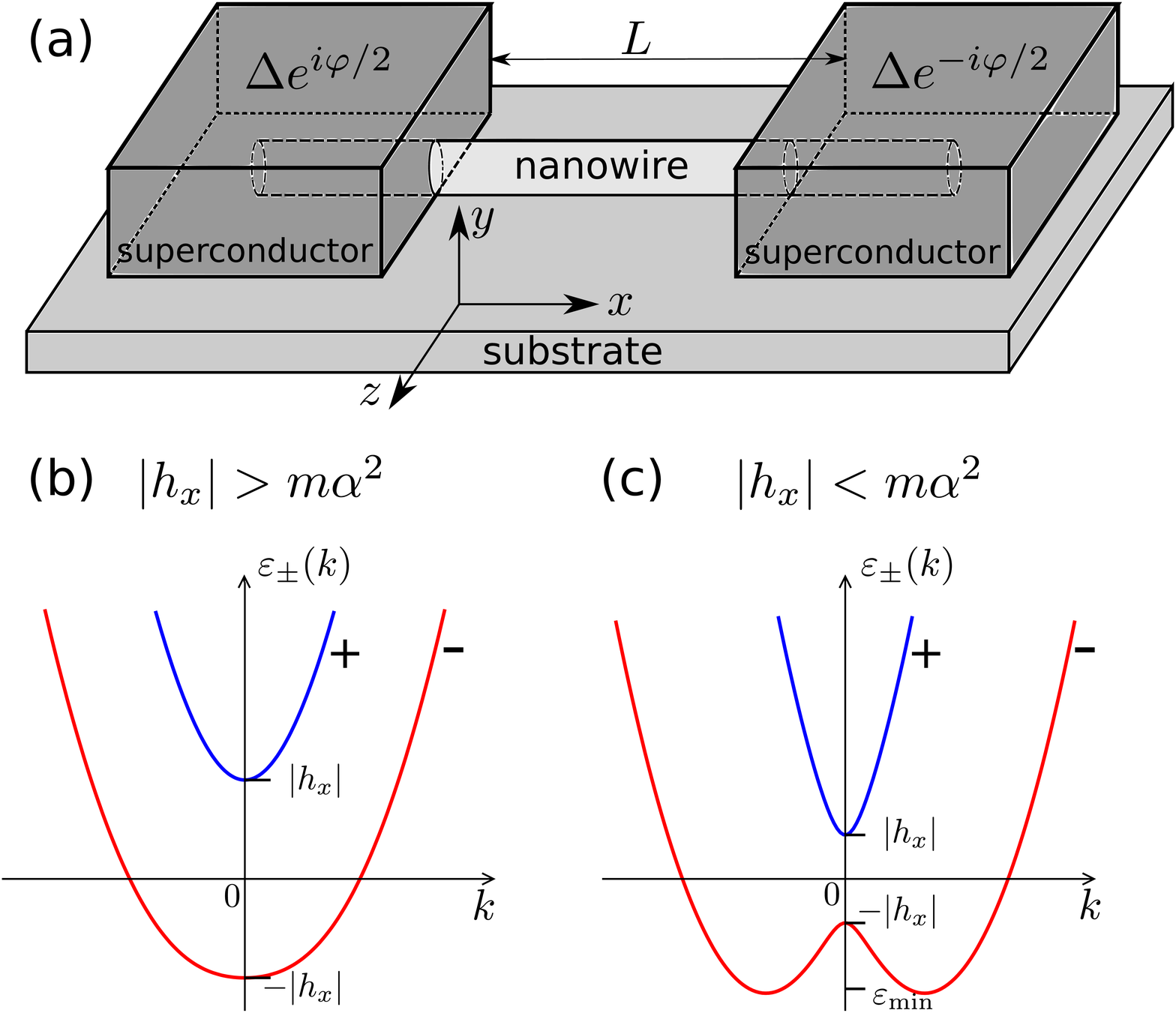}
 \caption{(a) The schematic setup involving a nanowire (light gray) placed in proximity with two superconductors (dark gray). The  effective magnetic field of Rashba SOC is along the $z$-axis, i.e., in the plane of the substrate and perpendicular to the nanowire.  (b), (c) The normal-state single-particle spectrum of the infinite nanowire described by the Hamiltonian~(\ref{BdG-nanowire}) with $h_z=0$ and with $|h_x| > m\alpha^2$ (b) and $|h_x| < m\alpha^2$ (c). }\label{Fig-nanowire}
\end{figure}

 In this paper, we show that the magnitude of the anomalous Josephson current allows one to identify whether a Josephson junction based on a nanowire with Rashba SOC and a Zeeman field is in the topologically nontrivial regime. In particular, we  consider  a setup shown schematically in Fig.~\ref{Fig-nanowire}(a), where two bulk superconductors  induce pairing gaps  in the left and right parts of the  nanowire via the proximity effect, while its middle part  is in the normal state. 
  We assume that the Zeeman field is present everywhere in the nanowire, including its superconducting parts. We find that, in a short junction, similarly to the QSHI system,~\cite{Dolcini2015} the dominant contribution to the AJE comes from the superconducting parts and increases very strongly when those parts are tuned into the topologically nontrivial phase. In particular, in that regime, the anomalous current $I_{\rm an}$ can become comparable to the critical current $I_c$. 
  Note that an AJE has been recently observed in junctions based on InSb nanowires.~\cite{Szombati2015}
 
 {
 The outline of the paper is as follows. 
 We introduce our model and discuss the condition for a uniform nanowire to be in the topologically nontrivial phase in  Sec.~\ref{Sec_model}. In Sec~\ref{Sec_short}, we study in detail the anomalous current in junctions with a short normal part.  
 In particular, we calculate the anomalous current flowing in such junctions in various regimes, check how it is affected by a normal part with a finite transmission probability, and discuss the current-phase relation. 
 In Sec.~\ref{Sec_long}, we consider junctions with a normal part of arbitrary length. We conclude in Sec.~\ref{Sec_concl}. 
 }
 
\section{Model}\label{Sec_model}
 
The nanowire  along the $x$-direction can be described by the Bogoliubov-de Gennes (BdG) Hamiltonian~\cite{Lutchyn2010,Oreg2010}
\begin{equation}\label{BdG-nanowire}
 \hat{H} = \left(\frac{\hat{{p}}_x^2}{2m}-\mu -\alpha \hat{p}_x \sigma_z \right) \tau_z - \textbf{h}\cdot{\boldsymbol \sigma}  -\Delta(x)\tau_+ -\Delta^*(x)\tau_-\,.
\end{equation}
Here $\mu$ is the chemical potential, $\alpha$ is the SOC strength, $\textbf{h} = (h_x,0,h_z)$ is the Zeeman field,  $\hbar = 1$,  the externally induced pairing potential is given by
\begin{equation}
 \Delta(x) = \Delta e^{i\varphi/2} \theta(-x-L/2) + \Delta e^{-i\varphi/2}\theta(x-L/2)\,,
\end{equation}
and $\tau_\pm = (\tau_x \pm i\tau_y)/2$. 
The Pauli matrices $\sigma_{x,y,z}$ and $\tau_{x,y,z}$ act in the spin and particle-hole spaces, respectively.

At $h_z=0$, the normal-state spectrum of an infinite nanowire is given by two nondegenerate bands,
\begin{equation}\label{sp-spectrum}
 \varepsilon_\pm(k)-\mu = \frac{k^2}{2m}-\mu \pm \sqrt{\alpha^2 k^2 + h_x^2}\,,
\end{equation}
and is shown in Figs.~\ref{Fig-nanowire}(b) and \ref{Fig-nanowire}(c) for  $|h_x| > m\alpha^2$ (b) and for $|h_x| < m\alpha^2$ (c). 
Further on, we label the quantities associated with the two bands by the respective signs, $\pm$. 
Note that for $|h_x|<m\alpha^2$, there exists a regime $\epsilon_{\rm min}<\mu<-|h_x|$, where the lower band has four Fermi points. In that case, we label the quantities pertaining to the smaller Fermi momentum by the superscript $+$.
When the chemical potential lies within the helical gap $-|h_x| <\mu < |h_x|$, spin is correlated with the direction of motion of an electron.~\cite{Streda2003}  Thus, for a sufficiently small $\Delta$, superconductivity in a uniform nanowire at $h_z=0$ is effectively spinless or helical when $|h_x|>|\mu|$. 

 For general values of $h_z$ and $\Delta$, the  uniform nanowire is in the topologically nontrivial phase provided the following conditions are met,  see Appendix~\ref{Sec_appendix_topol}:
\begin{equation}\label{topological_condition_general}
\left\{
\begin{array}{l}
h=\sqrt{h_x^2+h_z^2} > h_c = \sqrt{\mu^2+\Delta^2} \,,\\
|h_z| < \Delta\,.
\end{array}
\right.
\end{equation} 
The first inequality is the generalization of the well known topological condition $|h_x|>h_c$ at $h_z=0$,~\cite{Oreg2010,Lutchyn2010} while the second condition ensures that the spectrum remains gapped.~\cite{Romito2012, Rex2014}

Except Secs.~\ref{Sec_short_proj} and \ref{Sec_short_analytic}, where we provide  analytic expressions for the anomalous current  in short junctions at $|h_z|  \ll \Delta$, 
we use a tight-binding approximation to the Hamiltonian (\ref{BdG-nanowire}).~\cite{Cayao2015} The current is obtained by numerically finding its eigenvalues, which come in pairs $\pm E$, and then calculating the current as
\begin{equation}\label{current-numerics}
I = - e\sum' \frac{\partial E}{\partial \varphi}\,,
\end{equation}
where the sum is over the negative eigenvalues. 

Both the continuum eigenstates, $|E|>\Delta_{\rm min}$, where $\Delta_{\rm min}$ is the gap in the  quasiparticle spectrum, and the subgap Andreev bound states, $|E|<\Delta_{\rm min}$, contribute to this sum. In particular, one may distinguish two different contributions to the AJE. The magnetic field in the leads yields an effect of the order $I_{\rm an}/I_c \propto h_z/\Delta$ that can be ascribed to the continuum states. This effect is dominant in short junctions. The magnetic field in the junction yields an effect of the order $I_{\rm an}/I_c \propto h_z/E_{\rm Th}$, where $E_{\rm Th} \sim v_{F}/L$ is the Thouless energy and $v_{F}$ is the Fermi velocity. This effect can be ascribed to the bound states and is dominant in long junctions.



\section{Short junctions}\label{Sec_short}

We first consider a  short junction, $L\ll v_{F}^{\pm}/\Delta$, where $v_F^\pm$ are the Fermi velocities in the $\pm$-bands. 

In the limit
$L\to0$, the junction at $\varphi=0$ is equivalent to a uniform nanowire with a constant  $\Delta$. When $h_z=0$, such a nanowire does not carry a current  because of  the combined symmetry of $\sigma_z \hat{T}$, where $\hat{T}$ is the time-reversal operator.~\cite{Liu2010-PRB1} By contrast,
 when $\alpha$, $h_x$, and $h_z$ are all nonzero, one finds that the state with a constant $\Delta$ is not an energy minimum of the isolated nanowire, but an excited state with a finite supercurrent. Similarly to QSHI edge states~\cite{Dolcini2015} and 2D surface superconductors,~\cite{Dimitrova2007} the ground state would correspond to a modulated order parameter, $\Delta(x) \propto \exp[iqx]$, where $q\propto h_z$. As the proximity effect imposes a constant order parameter, the system cannot relax to its ground state and, therefore, displays an anomalous Josephson effect.

  In Secs.~\ref{Sec_short_proj} and \ref{Sec_short_analytic}, we discuss in detail the effect of a weak field along the $z$-direction, $|h_z|\ll\Delta$, in junctions with $\varphi=0$ and calculate 
 the linear response of the anomalous current to such a field. In Sec.~\ref{Sec_anisotropy}, we extend those results to larger values of $h_z$ by considering a general direction of a magnetic field ${\bf h} = h(\cos\theta, 0, \sin\theta)$. In Sec~\ref{Sec_barrier}, we demonstrate that the effect in short junctions in the topological phase - the magnitude of $I_{\rm an}$ in comparison to $I_c$ - is robust against inducing an electrostatic barrier in the normal part. In Sec.~\ref{Sec_cpr}, we  lift the condition $\varphi=0$ and discuss the  current-phase relation in short junctions. 
 
 \subsection{Projection into helical bands}\label{Sec_short_proj}
 
To understand the magnitude of the  effect and the roles of the helical bands  at $|h_z| \ll \Delta$, we first consider the limit where the superconducting correlations are mainly intra-band, which is the case when
\begin{equation}\label{projection_limit}
\Delta \ll 2\sqrt{(\alpha k_{F}^{\pm})^2 + h_x^2}\,,
\end{equation}
where $k_{F}^{\pm}$ are the Fermi momenta in the $\pm$-bands.
In that case, one can project the Hamiltonian~(\ref{BdG-nanowire})  into the helical basis. 
 To this end, we first rewrite the Hamiltonian~(\ref{BdG-nanowire}) in that basis, obtaining Eq.~(\ref{BdG_uniform_helical}) of Appendix~\ref{Sec_appendix_topol}. Ignoring the terms that mix the helical bands, i.e., keeping only the terms leading to first-order corrections in eigenenergies, we find that the projected Hamiltonian reads
\begin{equation}\label{Hamiltonian_effective}
 \hat{H} \approx \left(\xi_k - \sigma_z \sqrt{(\alpha k)^2 + h_x^2}\right)\tau_z - h_{\rm eff}(k)\sigma_z - \Delta_{\rm eff}(k) \tau_x\,.
\end{equation}
  Here  $\xi_k = k^2/(2m)- \mu$ and the effective field and pairing potential are given as
\begin{equation}\label{effective}
 \frac{h_{\rm eff}(k)}{h_z} = \frac{\Delta_{\rm eff}(k)}{\Delta} = \frac{\alpha k}{\sqrt{(\alpha k)^2 + h_x^2}}\,.
\end{equation}
 This is an effective Hamiltonian that describes electrons in two uncoupled $\pm$-bands with $k$-dependent  
 magnetic field $\mp h_{\rm eff}(k)$ and pairing gap $\Delta_{\rm eff}(k)$. 
 When $h_x\neq 0$, these  quantities are reduced with respect to their bare values because spins are no longer aligned along the $z$-direction.

In the topologically nontrivial phase, $|h_x|>\sqrt{\Delta^2+\mu^2}$,  only the  lower band contributes to the superconducting properties. 
Upon linearization of the normal-state spectrum, the system maps
onto the model of Ref.~\onlinecite{Dolcini2015} for the helical edge states,
and one finds the anomalous current
\begin{equation}\label{I_proj_top}
 I_{\rm an} = \frac{e}{\pi} h_{\text{eff}}^{-}\,,
\end{equation}
where $h_{\text{eff}}^{-}=h_{\rm eff}(k_F^-)$.
Since the spin direction  is tilted by the magnetic field $h_x$, the effect is reduced compared to the QSHI case. The maximal current $I_{\rm an}=I_h$, where
 $ I_h =  eh_z/\pi$, is reached only in the limit of large $m\alpha^2/|h_x|$. Note that the critical current, $I_c=e\Delta_{\rm eff}^-/2$, is reduced by the same factor because $\Delta_{\rm eff}^-/\Delta=h_{\text{eff}}^{-}/h_z$ with $\Delta_{\rm eff}^-=\Delta_{\rm eff}(k_F^-)$. 

For $\sqrt{\Delta^2+\mu^2} > |h_x|$, we  can map the projected Hamiltonian onto two copies of the QSHI model. As the effective fields have opposite signs in the two bands, this results in the anomalous current
\begin{equation}\label{I_proj_nontop}
 I_{\rm an} = \frac{e}{\pi} h_{\text{eff}}^{-} - \frac{e}{\pi} h_{\text{eff}}^{+}\,,
\end{equation}
 where $h_{\text{eff}}^{+}=h_{\rm eff}(k_F^+)$. Thus, in the topologically trivial regime, the AJE is strongly reduced due to the competition between the two bands. In particular, it exists only if the effective fields are different, which requires $h_x\neq 0$.

The projection onto the helical bands requires that the latter are sufficiently split in energy. This yields the condition (\ref{projection_limit}). Furthermore, 
the chemical potential should be sufficiently far away from the band thresholds, $\left|\mu - |h_x|\right| \gg \Delta$, so the normal spectrum can be linearized when mapping onto the QSHI model.
 
\subsection{Exact analytic calculations at $|h_z| \ll \Delta$}\label{Sec_short_analytic}
  
To generalize Eqs.~(\ref{I_proj_top}) and (\ref{I_proj_nontop}) to arbitrary values of $\Delta$, we calculate the supercurrent at $L=0$ and $\varphi=0$ as  $I=-2e\partial f/\partial q|_{q=0}$, where $f$ is the free-energy density  in the presence of a modulated order parameter $\Delta\exp[iqx]$. This density is given by $f = -(1/2) \int (dk/2\pi) (\tilde{E}_{k+}+\tilde{E}_{k-})$, where $\tilde{E}_{k\pm}$ are the positive eigenvalues of the BdG Hamiltonian. Evaluating  them perturbatively in $q$ and $h_z$, we find exactly the linear response of the supercurrent to  a finite $h_z$ as (see Appendix~\ref{Sec_appendix_exact})
\begin{equation}\label{I_exact}
 I = I_h \alpha h_x^2 \int \limits_{-\infty}^{\infty} dk \frac{(E_{k+} + E_{k-})^2 - 4(\xi_k^2+\Delta^2)}{E_{k+}E_{k-} (E_{k+} + E_{k-})^3}\,.
\end{equation}
Here, $E_{k\pm}$ are the eigenvalues at $q,h_z=0$ given by 
\begin{equation}\label{E_k_spectrum}
 E_{k\pm}^2 = \xi_k^2 + (\alpha k)^2 + h_x^2 + \Delta^2 \pm 2\sqrt{\xi_k^2(\alpha k)^2 + \xi_k^2 h_x^2 + h_x^2\Delta^2}\,.
\end{equation}
The previous results, Eqs.~(\ref{I_proj_top}) and (\ref{I_proj_nontop}), can be recovered by taking the limit $\Delta\to0$  in this expression.

\begin{figure}[t]
 \includegraphics[width=\columnwidth]{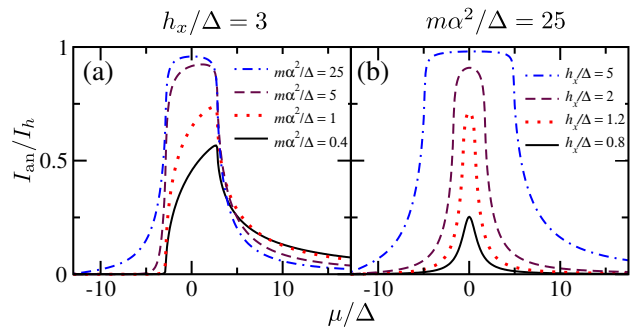}
 \caption{The anomalous Josephson current $I_{\rm an}$ in units of the maximal anomalous current $I_h = eh_z/\pi$ as a function of $\mu/\Delta$ in a short junction  in the presence of a small $h_z$. The results are shown for $h_x/\Delta=3$ and several values of $m\alpha^2/\Delta$ (a) and for $m\alpha^2/\Delta=25$ and several values of $h_x/\Delta$ (b).  }\label{Fig-short}
\end{figure}

The anomalous current of a short junction, obtained from  Eq.~(\ref{I_exact}), is shown  in Fig.~\ref{Fig-short} as a function of $\mu/\Delta$.
This figure conveys our main message:  The AJE is large in the topologically nontrivial phase determined by $|\mu| < \mu_c = \sqrt{h_x^2 - \Delta^2}$ [see Eq.~(\ref{topological_condition_general}) at $h_z=0$], whereas it rapidly decays at larger $|\mu|$.

The different shape of the curves for $m\alpha^2 < |h_x|$ and $m\alpha^2 > |h_x|$  is well explained by the projection into the helical bands. In the limit $m\alpha^2/|h_x| \gg 1$, we find $h_{\rm eff}^- \approx h_z$, such that  Eq.~(\ref{I_proj_top}) predicts a plateau $I_{\rm an} = I_h$ in the topological regime, which is seen at  large $|h_x|/\Delta$ (see the dash-dotted lines in both panels). By contrast,
in the limit $m\alpha^2/|h_x| \ll 1$, we find $h^-_{\rm eff}/h_z  \approx \alpha k_F^-/|h_x|$, which, in the helical gap, increases monotonically starting from $0$ and, therefore, gives an asymmetric dependence of $I_{\rm an}$ on $\mu$ with a peak at $\mu \sim \mu_c$ [see the solid and dotted curves in Fig.~\ref{Fig-short}(a)]. For large chemical potentials, $\mu \gg |h_x|, \, m\alpha^2$, the current decays as  $I_{\rm an}/I_h \approx h_x^2/(\sqrt{2m\alpha^2}\mu^{3/2})$. 
 
 Figure~\ref{Fig-short}(b) shows the effect of increasing the ratio $\Delta/h_x$. The range in $\mu$, where the system is in the topologically nontrivial regime, becomes narrower and completely disappears at 
 $\Delta=|h_x|$. Furthermore, the value of the anomalous current in the topological regime decreases  because of an averaging over an energy window of width $\Delta$. This averaging suppresses the effect compared to the result~(\ref{I_proj_top}), as the spin projection of the two bands varies with energy. Note that a residual enhancement of $I_{\rm an}$ around $\mu=0$ remains in the topologically trivial regime because the difference between the effective fields decreases with $|\mu|$.

\subsection{Magnetic-field anisotropy}\label{Sec_anisotropy}

\begin{figure}
 \includegraphics[width = \columnwidth]{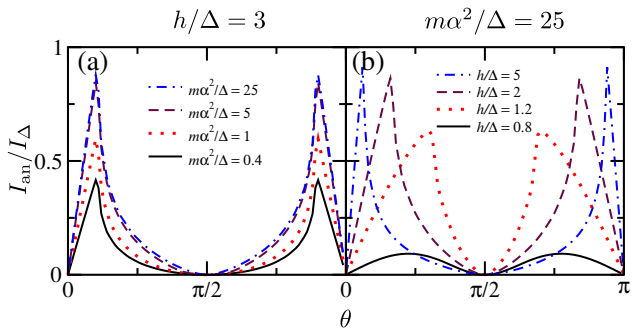}
 \caption{  The anomalous current $I_{\rm an}$  in a short junction  at $\mu=0$ in units of $I_\Delta = e\Delta/\pi$ as a function of the angle $\theta$ that parametrizes the Zeeman field, ${\bf h} = h(\cos\theta, 0, \sin\theta)$. The results are shown for $h/\Delta = 3$ and several values of $m\alpha^2/\Delta$ (a) and for $m\alpha^2/\Delta = 25$ and several values of $h/\Delta$ (b). Note the cusp at $h_z=h\sin\theta_c=\Delta$.}\label{Fig-thetadep-short}
\end{figure}

Here we extend our considerations of short junctions to larger fields by numerically studying the AJE for 
a general direction of the Zeeman field, ${\bf h} = h(\cos\theta, 0, \sin\theta)$. 
In Fig.~\ref{Fig-thetadep-short}, we show the anomalous current in a short junction as a function of $\theta$ at $\mu=0$ for the same values of $m\alpha^2$ and $h$ as in Fig.~\ref{Fig-short} (with $h_x$  replaced by $h$). When $h>\Delta$, the anomalous current increases with increasing $\theta$ in the topologically nontrivial phase until we reach the condition $h_z=h\sin\theta_c=\Delta$, where the system becomes gapless [see the topological condition~(\ref{topological_condition_general})]. Note that the anomalous current reaches values of the order of the critical current, $I_{\rm an}\sim I_\Delta=e\Delta/\pi$, i.e., the effect is large. At $\theta=\theta_c$ the anomalous current has  a cusp, while for $\theta >\theta_c$ it decays down to zero at  $\theta=\pi/2$, where $h_x=0$. The results are symmetric around $\theta=\pi/2$ since the transformation $h_x \rightarrow -h_x$  is equivalent to the  transformation of the BdG Hamiltonian $\hat{H} \rightarrow \sigma_z \hat{H}\sigma_z$, which does not change the eigenvalues. 
When $h<\Delta$, the system is topologically trivial at all angles and, as can be seen 
in panel (b) of Fig.~\ref{Fig-thetadep-short}, the anomalous current is reduced and its angular dependence is smooth.

\subsection{Junctions with imperfect transmission}\label{Sec_barrier}

\begin{figure}[t]
\includegraphics[width=\columnwidth]{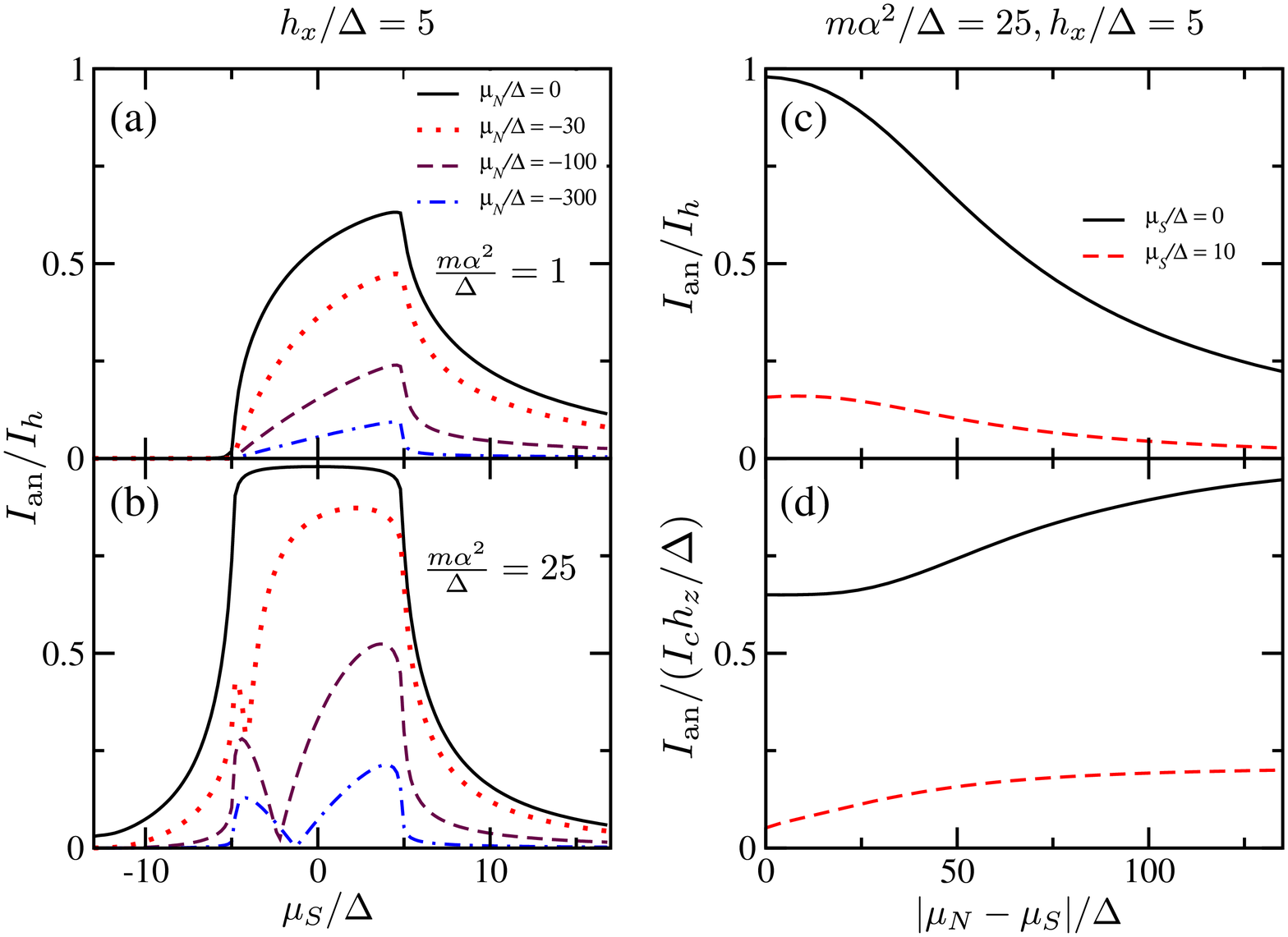}
 \caption{Left panels: The anomalous current $I_{\rm an}/I_h$ at $h_x/\Delta= 5$ and $h_z/\Delta=0.1$ as a function of $\mu_S/\Delta$ for $m\alpha^2/\Delta =1$ (a) and $m\alpha^2/\Delta = 25$ (b), shown for several values of $\mu_N/\Delta$. Right panels: The anomalous current $I_{\rm an}$ as a function of the barrier height $|\mu_N-\mu_S|/\Delta$ for $\mu_S/\Delta=0$ (solid lines) and $\mu_S/\Delta=10$ (dashed lines), using the same other parameters as in panel (b). The current is shown in units of $I_h$ (c) and in units of $I_c h_z/\Delta$ (d), where the critical current $I_c$ is calculated at $h_z=0$. 
}\label{Fig-barrier}
\end{figure}

In an experimental setup with multiple gate electrodes, one can control separately the electrostatic potentials of the normal and superconducting parts of the nanowire. This is  described by a position-dependent 
\lq\lq chemical potential\rq\rq, $\mu(x) = \mu_S\theta(|x|-L/2) + \mu_N\theta(L/2-|x|)$, in Eq.~(\ref{BdG-nanowire}). To study the effects of $\mu_N < \mu_S$  in the tight-binding formalism, we take the hopping energy to be $t/\Delta = 25$ and assume only one site in the normal part, which corresponds to a $\delta$-function barrier in the continuous model. 
Figures~\ref{Fig-barrier}(a) and \ref{Fig-barrier}(b) show the dependence of the anomalous current on $\mu_S/\Delta$ for several values of $\mu_N \leq 0$. When $|h_x|>m\alpha^2$ [see Fig.~\ref{Fig-barrier}(a)], we observe a smooth suppression of the anomalous current when decreasing $\mu_N$, which corresponds to increasing the height of the  barrier. The enhancement of  $I_{\rm an}$  in the topologically nontrivial regime remains for all values of $\mu_N$. The case 
$|h_x|<m\alpha^2$ [see Fig.~\ref{Fig-barrier}(b)] is more complicated.  In addition to an overall suppression of the anomalous current, resonances appear in the topologically nontrivial regime. They are signatures of Fano dips in the transmission probability, which exist due to the presence of the second band in the junction area, while the superconducting leads are topological.~\cite{Cayao2015}

\begin{figure}[t]
 \includegraphics[width=\columnwidth]{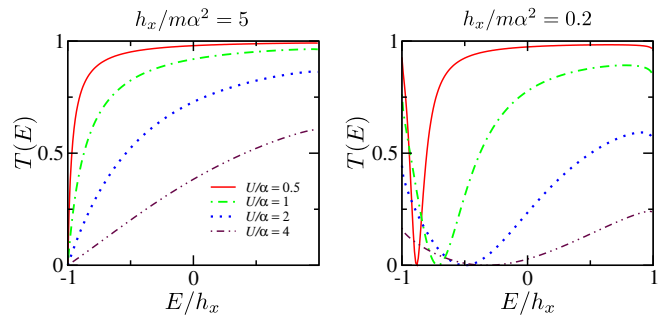}
 \caption{ Normal-state transmission $T$ through a Dirac potential as a function of the energy $E$ of the incident particle. The results are shown in the helical gap ($|E| < |h_x|$) for several values of the barrier height $U$  for $|h_x| > m\alpha^2$ (left panel) and for $|h_x| < m\alpha^2$ (right panel). }\label{Fig-transmission}
\end{figure}

In Fig.~\ref{Fig-transmission}, we plot the normal-state transmission probability $T$ as a function of energy $E$ in the helical gap $-|h_x|<E<|h_x|$ for a nanowire with the Dirac-potential barrier $V(x) = U\delta(x)$, see Appendix~\ref{Sec_appendix_transmission} for details of analytic calculations. 
The transmission is a universal function of $E/m\alpha^2$, $h_x/m\alpha^2$, and $U/\alpha$. We observe that, for $|h_x| > m\alpha^2$ (left panel), $T(E)$ increases smoothly starting from $T = 0$ at $E=-|h_x|$, as the system is insulating below that energy. In the opposite case (right panel), the system is conducting at $\epsilon_{\rm min} < E < -|h_x|$, so $T(-|h_x|)\ne 0$. Furthermore, we observe resonances (Fano dips) in the transmission.
They can be attributed to the formation of a quasi-bound state attached to the inverted parabolic part of the spectrum near $k=0$, cf. Fig.~\ref{Fig-nanowire}(c), for which the barrier acts as a quantum well.

Away from the resonances  of Fig.~\ref{Fig-barrier}(b),  an enhancement of the anomalous current in the topologically nontrivial regime remains. 
This is further illustrated in Fig.~\ref{Fig-barrier}(c), where we plot $I_{\rm an}/I_h$ vs $|\mu_S-\mu_N|$ for the topologically nontrivial  ($\mu_S=0$) and trivial ($\mu_S=10\Delta$) phases. We further note that, while the anomalous current decreases in the presence of a barrier, it decreases more slowly than the critical current. This is shown in Fig.~\ref{Fig-barrier}(d), where  the ratio $I_{\rm an}/I_c$ increases with $|\mu_S-\mu_N|$ (in both regimes).  Thus, we conclude that a large anomalous Josephson current is a robust signature  of  the topologically nontrivial phase.

In the topologically nontrivial regime, this increase of $I_{\rm an}/I_c$ is well described by the analytical results obtained from an extension of the formalism used in Ref.~\onlinecite{Dolcini2015} to a system with a finite transmission $T$. In particular, one finds
\begin{equation}\label{I_an_barrier}
I_{\rm an}=\frac e\pi h_{\rm eff}^-\sqrt{\frac T{1-T}}\arctan\sqrt{\frac{1-T}T}
\end{equation}
and $I_c=(e\Delta_{\rm eff}^-/2)\sqrt T$. Namely, $I_{\rm an}/I_c$ increases from $I_{\rm an}/I_c=2/\pi$ at $T=1$ to $I_{\rm an}/I_c=1$ at $T\ll1$  (in units of $h_z/\Delta$). This result applies sufficiently far away from the Fano dips, where the energy dependence of the transmission can be neglected.

\begin{figure}[t]
 \includegraphics[width=\columnwidth]{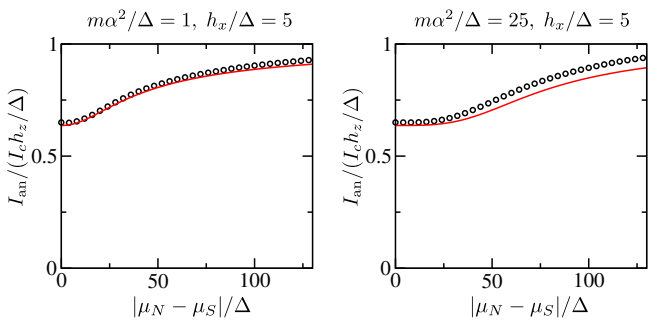}
 \caption{ Ratio between the anomalous and critical current, $I_{\rm an}/(I_ch_z/\Delta)$, as a function of the height of the electrostatic barrier, in units of $(\mu_S-\mu_N)/\Delta$, at $\mu_S=0$. The black dots are the numerical results from a tight-binding Hamiltonian with bandwidth $t/\Delta=25$. The red curves are the result of the helical edge model, Eq.~(\ref{I_an_barrier}), with the transmission $T=T(E=0)$ obtained from the continuous model. Left and right panels correspond to two different sets of parameters $h_x/\Delta$ and $m\alpha^2/\Delta$. The black dots in the right panel correspond to the solid line in Fig.~\ref{Fig-barrier}(d). }\label{Fig-comparison}
\end{figure}

 In Fig.~\ref{Fig-comparison}, we compare the numerical results obtained from the tight-binding model (Fig.~\ref{Fig-barrier}) with those obtained from the helical edge model at finite transmission $T=T(E=0)$, see Eq.~(\ref{I_an_barrier}), where we identified $U/\alpha$ with $(\mu_S-\mu_N)/\sqrt{m\alpha^2 t}$. Note that we choose $\mu_S=0$ so that Fano dips are separated from the relevant energy for the Josephson coupling when $\Delta$ is sufficiently small. The agreement is better for the parameters of the left panel because they are closer to the assumption of a wide band that was assumed in the continuum model, and because the energy dependence of the transmission cannot be neglected for the parameters of the 
right panel, which yield Fano dips.

\subsection{Current-phase relation}\label{Sec_cpr}

\begin{figure}
  \includegraphics[width=\columnwidth]{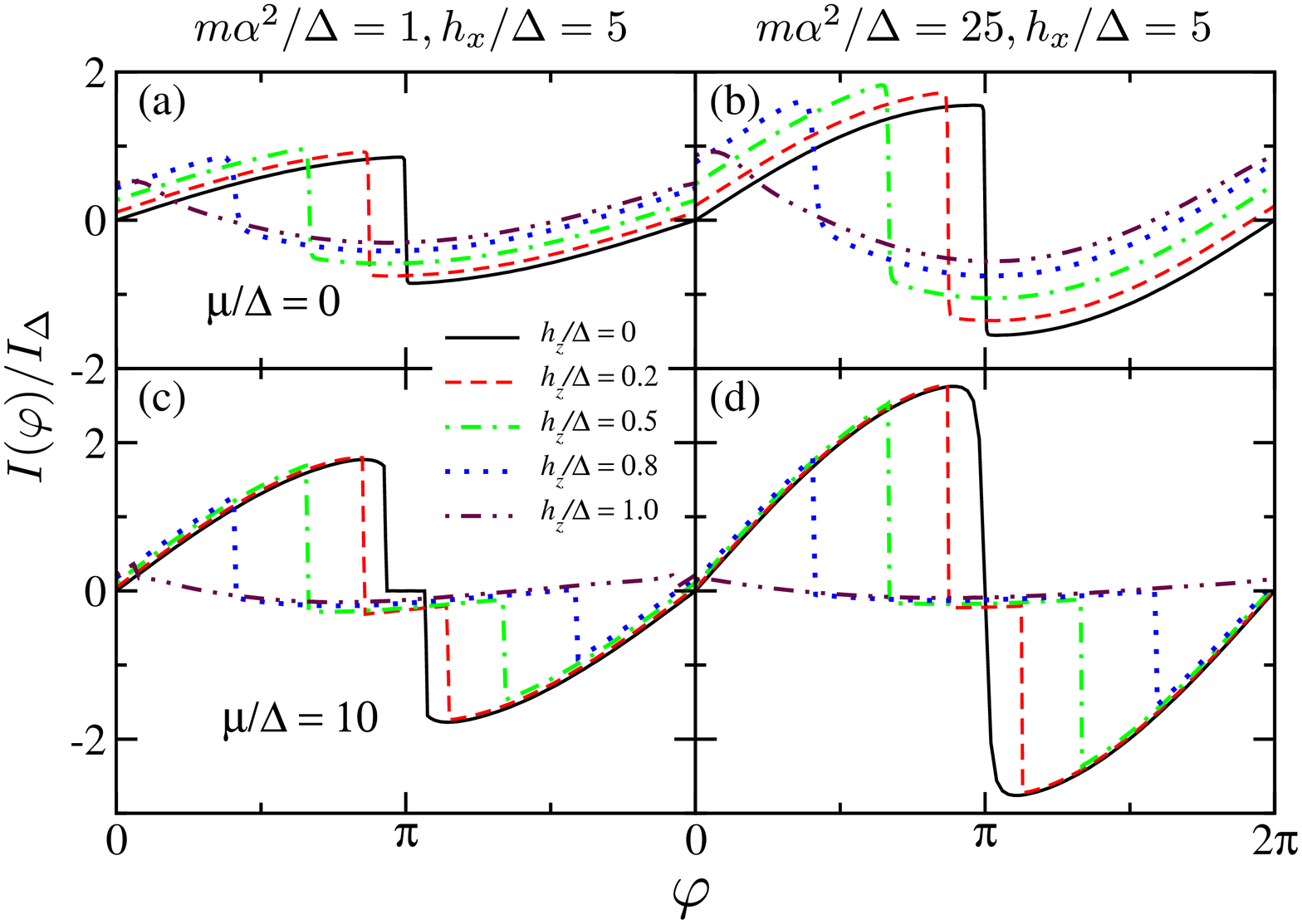}
  \caption{ The supercurrent $I(\varphi)$ in units of $I_\Delta$ flowing in a short junction ($L=0$) as a function of the phase difference $\varphi$. The top panels correspond to the topologically nontrivial regime ($\mu/\Delta = 0$), whereas the bottom panels correspond to the topologically trivial regime ($\mu/\Delta = 10$). The results are shown for several values of $h_z \le \Delta$ for $m\alpha^2 < |h_x|$ (left panels) and $m\alpha^2 > |h_x|$ (right panels).}\label{Fig-cpr}
 \end{figure}

 In this section, we briefly discuss the  current-phase relation (CPR) in a short junction ($L\rightarrow 0$), concentrating on the dependence on $h_z$. The numerical results are shown in Fig.~\ref{Fig-cpr}. We find that many  features of the current-phase relation are well explained by the projection into the helical bands with the mapping onto the QSHI model, see  Eqs.~\eqref{Hamiltonian_effective} and \eqref{effective}.
 
When the nanowire is in the topologically nontrivial phase, the system is described by the QSHI model with effective parameters $h_{\rm eff}^-=h_{\rm eff}(k_F^-)$ and $\Delta_{\rm eff}^-=\Delta_{\rm eff}(k_F^-)$, see Eq.~\eqref{effective}. The zero-temperature CPR is given as~\cite{Dolcini2015}
\begin{equation}
I_J(h_z,\varphi)= \frac{e}{\pi}h_{\rm eff}^--\frac e2\Delta_{\rm eff}^-\sin\frac\varphi2\,{\rm sign}\left(\sin\frac{\varphi-\varphi^*}2\right).
\end{equation}
It displays a single jump associated with the zero-energy crossing of the Andreev bound state formed in the junction at phase $\varphi^*=\pi-2\arcsin(h_z/\Delta)$.

When the nanowire is in the topologically trivial phase (see bottom panels of Fig.~\ref{Fig-cpr}), the projection into the helical bands predicts two jumps in the CPR at phases $\varphi^*_\pm=\pi\pm2\arcsin(h_z/\Delta)$. The two jumps move in the opposite directions with increasing $h_z$ since the effective magnetic field has different signs in the two bands. Thus, the differences in the CPR between the topologically trivial and nontrivial regimes are more pronounced at finite $h_z$. 

Residual interband pairing in the topologically trivial regime, which is not captured by the projection, hybridizes the two Andreev bound states formed in the junction. At $h_z=0$, depending on parameters, this either leads to the opening of a gap at zero energy, resulting in a rounding of the jump as can be seen in Fig.~\ref{Fig-cpr}(d) for $|h_x|\ll m\alpha^2$, or in a small shift of the zero-energy crossings to phases $\pi\pm\delta\varphi$ (see Ref.~\onlinecite{Lutchyn2010}), resulting in two jumps in the CPR as can be seen in Fig.~\ref{Fig-cpr}(c) for $|h_x|\gg m\alpha^2$. The effects are less visible at finite $h_z$.

\section{Long junctions}\label{Sec_long}

\begin{figure}[t]
 \includegraphics[width=\columnwidth]{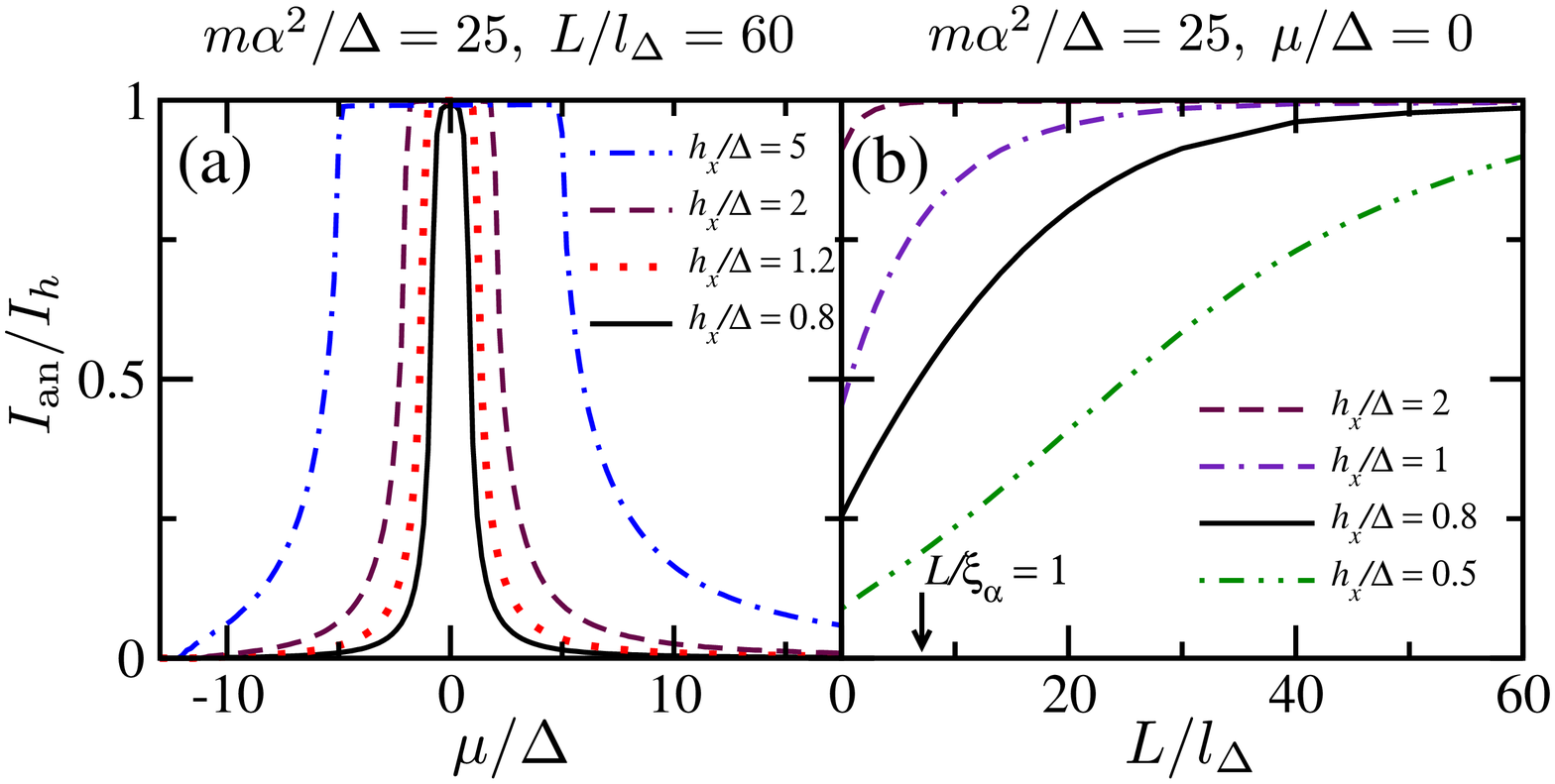}
 \caption{The  anomalous current $I_{\rm an}/I_h$ at $h_z/\Delta=  0.01$ for junctions with finite length $L/l_\Delta$.  (a) $I_{\rm an}/I_h$ vs $\mu/\Delta$ for a  junction with $L/l_\Delta=60$ for the same parameters as in Fig.~\ref{Fig-short}(b). In the topological regime, this length corresponds to $L/\xi_{\alpha} \approx 8.5$. (b) $I_{\rm an}/I_h$ vs $L/l_\Delta$ calculated at $\mu=0$ for $m\alpha^2/\Delta=25$ and  several values of $h_x/\Delta$. The value of $L/l_\Delta$ for which $L/\xi_\alpha\approx 1$ is shown by a vertical arrow. 
  }\label{Fig-long}
\end{figure}

Let us now turn to junctions of arbitrary length $L$. Normally, short and long junctions are distinguished by the ratio $L/\xi$, where $\xi$ is the superconducting coherence length. Here $\xi$ strongly varies as a function of $\alpha$, $\mu$, and $h_x$. Thus, we use $l_\Delta = 1/\sqrt{2m\Delta}$ as the reference length scale. In the tight-binding model, the length of the junction is then given as  $L= (N/\sqrt{t/\Delta})l_\Delta$, where $t$ is the hopping energy and $N$ the number of sites in the normal part.

In long junctions, the main contribution to the anomalous current is due to the phase accumulated in the junction area, which is of the order of $h_{\rm eff}^\pm/E_{\rm Th}^\pm$, where the Thouless energy $E_{\rm Th}^\pm=v_F^\pm/L\ll\Delta$. As a result the anomalous current is mainly sensitive to the properties of the normal part rather than the properties of the superconducting leads.

Thus, the anomalous effect is expected to be large when the normal part is helical.
In Fig.~\ref{Fig-long}(a), we show  $I_{\rm an}/I_h$ vs $\mu/\Delta$ for the same parameters as in Fig.~\ref{Fig-short}(b), but for a junction with $L/l_\Delta = 60$, which corresponds to $L/\xi_\alpha\approx 8.5$, where $\xi_\alpha=\alpha/\Delta$ is the characteristic scale for superconducting correlations in the helical regime. A plateau in the helical regime $|\mu|<|h_x|$, where the anomalous current takes the value $I_{\rm an}/I_h=h_{\rm eff}^-/h_z$, is clearly visible. To explain this, we note that, in the QSHI model,~\cite{Dolcini2015} the magnitude of the effect at small $h_z$ is universal, i.e., independent of $L$. Therefore, in the nanowire, it should be universal as well when the result (\ref{I_proj_top}) is applicable. 

In a long junction, $\Delta$ in the condition (\ref{projection_limit}) has to be replaced by $E_{\rm Th}$. At sufficiently small  $E_{\rm Th}$, Eq.~(\ref{I_proj_top}) then holds in the entire regime $|\mu|<|h_x|$. When the leads are topologically trivial  while the normal part is helical, the normal part acts as a barrier for the upper band, suppressing its contribution. 
As shown in Fig.~\ref{Fig-long}(b), this implies an increase of the anomalous current with increasing junction length in the regime  $h_x^2-\Delta^2<\mu^2<h_x^2$. Furthermore, the current also increases with junction length when the superconductors are in the topological regime. While in a short junction, averaging over an energy window of width $\Delta$ lead to a suppression of the effect, here the relevant scale is $E_{\rm Th}\ll\Delta$. Furthermore, it is worth noting that, while in short junctions Eqs.~(\ref{I_proj_top}) and (\ref{I_proj_nontop}) are applicable for $|h_z|<\Delta$, in long junctions they apply only for $|h_z| < E_{\rm Th}^\pm \ll \Delta$.

\begin{figure}
 \includegraphics[width = \columnwidth]{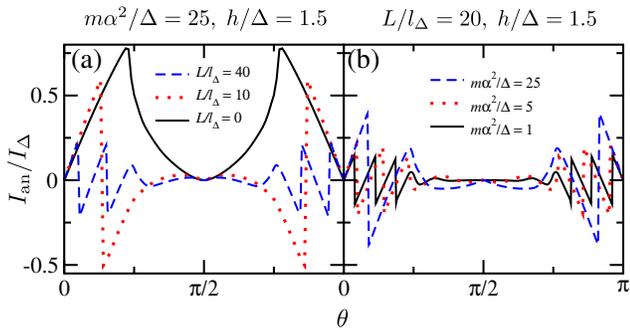}
 \caption{ The anomalous current $I_{\rm an}/I_\Delta$ vs $\theta$ in a long junction at $\mu=0$ and $h/\Delta = 1.5$. The results are shown for $m\alpha^2/\Delta = 25$ and several values of $L/l_\Delta$ (a) and for $L/l_\Delta = 20$ and several values of $m\alpha^2/\Delta$ (b). In panel (a), the result for a short junction ($L/l_\Delta=0$) is shown for comparison.}\label{Fig-thetadep-long}
\end{figure}

We now study larger values of $h_z$, assuming a general direction of magnetic field $\bf h = h(\cos\theta, 0, \sin\theta)$.
Figure~\ref{Fig-thetadep-long} shows the results for a long junction, where the contribution to the anomalous current is due to the phase accumulated within the normal part of the junction. When the superconducting leads are in the topologically nontrivial regime, the mapping to helical edge states predicts the current to be a function of $\varphi + \varphi_h$, where $\varphi_h = 2h_{\text{eff}}^-/E_{\text{Th}}^-$. The current-phase relation has a jump when this argument changes by $2\pi$ and one bound state crosses zero. As a consequence  the anomalous current displays a sawtooth behavior as a function of $h_z$ with the spacing between jumps being $\sim E_{\rm Th}^-$. The sawtooth behavior disappears at $\theta \sim \theta_c$, when the system becomes gapless. In Fig.~\ref{Fig-thetadep-long}(a), we plot the anomalous current $I_{\rm an}$ as a function of $\theta$ for junctions with several values of $L/l_\Delta$ while keeping all other parameters the same. We see that the distance between two jumps decreases 
with increasing $L$, 
corresponding to decreasing $E_{\text{Th}}^-$. In Fig.~\ref{Fig-thetadep-long}(b), we 
fix $L/l_\Delta$ and vary $E_{\text{Th}}^-$ by changing $m\alpha^2$. Likewise, we observe that this distance between jumps is shorter for a junction with smaller $\alpha$ and, therefore, with smaller $v_{F}^-$ and $E_{\text{Th}}^-$.

\section{Conclusion}\label{Sec_concl}

In conclusion, we have studied the anomalous supercurrent in Josephson junctions made with semiconducting nanowires. Such an effect is possible in the presence of a magnetic field with nonzero components both along the effective field of Rashba spin-orbit coupling and along the nanowire. We have demonstrated that this current can increase very strongly if various parts of the system are tuned into the helical or topologically nontrivial regimes. In particular, a large value of the anomalous current in the short-junction limit  indicates that the superconducting parts are topological. The results presented here could describe realistic junctions made with nanowires such as that of Ref.~\onlinecite{Szombati2015}, where $m\alpha^2 \sim \Delta$.

 {\sl Note added.} Recently, we became aware of Ref.~\onlinecite{Wu2016} where a similar conclusion about a large AJE being a signature of topological superconductivity was reached. In that reference, a junction with two nanowires and a semiconducting ring between them was studied numerically. 

\begin{acknowledgments}
We would like to thank Fabrizio Dolcini for very stimulating discussions in the early stages of this work. 
Furthermore, we acknowledge support by ANR through grants
ANR-11-JS04-003-01 and ANR-12-BS04-0016-03, and by
an EU-FP7 Marie Curie IRG.
\end{acknowledgments}

\begin{appendix}
 \section{Superconductivity induced in a uniform nanowire}\label{Sec_appendix_uniform}

 Here we first derive the conditions~(\ref{topological_condition_general}) for the superconductivity induced in a uniform nanowire to be topological. Then, we provide details on the analytical derivation of the anomalous current in a short junction [Eq.~(\ref{I_exact})].

\subsection{Topological criterion}\label{Sec_appendix_topol}

The Hamiltonian which describes the nanowire in the presence of a constant pairing potential $\Delta$ can be written as
$
 \hat{H} = (1/2)\sum_k \Gamma_k^\dagger {\mathcal{H}}_k \Gamma_k + {\rm const.}\,
$,
where $\Gamma_k = (c_{k\uparrow},  c_{k \downarrow}, c^\dagger_{-k \downarrow}, -c^\dagger_{-k\uparrow})^T$ combines electron creation and annihilation operators and 
\begin{equation}\label{BdG_uniform_0}
 {\mathcal{H}}_k = \xi_k\tau_z - \gamma_k \sigma_z  \tau_z - h_x\sigma_x - h_z\sigma_z - \Delta\tau_x
 \end{equation}
is the BdG Hamiltonian.  Here, $\xi_k = k^2/(2m) - \mu$ and $\gamma_k = \alpha k$.

Using the transformation ${\mathcal{H}}_k \rightarrow U_k^\dagger {\mathcal{H}}_k U_k$ with $U_k = \exp\left[-i(\theta_k/2)  \sigma_y\tau_z\right]$ and $\theta_k =\arccos(\gamma_k/\sqrt{\gamma_k^2 + h_x^2})$, we may write the BdG Hamiltonian (\ref{BdG_uniform_0}) in the helical basis as
\begin{widetext}
 \begin{equation}\label{BdG_uniform_helical}
 {\mathcal{H}}_k = \left(\xi_k - \sqrt{\gamma_k^2 + h_x^2}\;\sigma_z\right) \tau_z - h_z\left(\cos\theta_k \sigma_z - \sin\theta_k \sigma_x \tau_z\right)  - \Delta\left(\cos\theta_k \tau_x - \sin\theta_k \sigma_y \tau_y\right)\,.
 \end{equation}
 \end{widetext}
 Since $U_k^\dagger \Gamma_k = (a_{k,-}, a_{k,+}, a^\dagger_{-k,-}, a^\dagger_{-k,+})^T$, where $a^\dagger_{k,\pm}$ and $a_{k,\pm}$ are the creation  and annihilation operators of  electrons in $\pm$ helical bands,  the terms  $- \Delta\cos\theta_k \tau_x$ and $\Delta\sin\theta_k \sigma_y \tau_y$ describe, respectively, intra- and interband pairing.

Squaring twice Eq.~(\ref{BdG_uniform_helical}) after subtracting a constant term, we find that its eigenvalues solve the equation
\begin{equation}\label{BdG-eigenvalues-uniform}
 \left(E^2 - \bar{E}^2_{k+}\right) \left(E^2 - \bar{E}^2_{k-}\right) = 4\gamma_k^2{h}_z^2 + 8{\xi}_k \gamma_k {h}_z E\,,
\end{equation}
where
\begin{equation}\label{bar_E_kpm}
\bar{E}_{k\pm}^2 = {\xi}_k^2 + \gamma_k^2 + {h}^2 + \Delta^2 \pm 2\sqrt{{\xi}_k^2 \gamma_k^2 + {\xi}_k^2 {h}^2  + {h}^2\Delta^2}
\end{equation}
and 
\begin{equation}
h^2 = h_x^2 + h_z^2\,.
\end{equation}

Equation~(\ref{BdG-eigenvalues-uniform}) allows to formulate the criterion of the topological phase transition, which is characterized by zero-energy eigenvalues. 
Such an eigenvalue exists provided the equation
\begin{equation}
\bar{E}_{k+}^2 \bar{E}_{k-}^2 - 4\gamma_k^2 h_z^2 = (\xi_k^2 - \gamma_k^2 - h^2 + \Delta^2)^2 - 4\gamma_k^2 (h_z^2 - \Delta^2) = 0 
\label{eq-topo}
\end{equation}
is satisfied for some $k$.  
 When $|h_z| < \Delta$,  the solution exists if and only if $\gamma_k=0$ (i.e., $k=0$) and $h^2 = h_c^2 = \Delta^2 + \mu^2$.  When $h\ne h_c$ and $|h_z|<\Delta$, the Hamiltonian can be transformed continuously into the Hamiltonian with $h_x = h$ and $h_z=0$ without closing the gap. Using the known topological criterion for the transformed Hamiltonian, we find that, for $|h_z|<\Delta$, the system is in the nontrivial phase provided $h>h_c$. If $|h_z|>\Delta$,  Eq.~\eqref{eq-topo} always has a solution with some finite $k$, i.e., the phase is gapless. 

\subsection{Anomalous current in a short nanowire}\label{Sec_appendix_exact}

As we argued in Sec.~\ref{Sec_short}, the anomalous current in a short nanowire junction is identical to the supercurrent $I$ which flows in a superconducting nanowire with a spatially uniform pairing potential. 

To calculate this supercurrent, we may use the relation $I = -2e\partial f/\partial q|_{q\to 0}$, where $f$ is the free-energy density in the presence of a modulated order parameter $\Delta(x) = \Delta \exp({iqx})$. To this end, we derive the corresponding BdG Hamiltonian,
\begin{equation}\label{BdG-q}
 \tilde{\mathcal{H}}_k  = \left(\tilde{\xi}_k -\gamma_k \sigma_z \right) \tau_z - h_x \sigma_x - \tilde{h}_z\sigma_z  -\Delta \tau_x + \frac{kq}{2m},
 \end{equation}
where $\tilde{\xi}_k = \xi_k + q^2/(8m)$ and $\tilde{h}_z = h_z + \alpha q/2$. Using perturbation theory, we find that its eigenenergies are given by $E_{k\pm}+\delta E_{k\pm}$, where $E_{k\pm}$ are the eigenenergies at $h_z=0$ and $q=0$ [see Eq. (\ref{E_k_spectrum})] and
\begin{eqnarray}
 \delta E_{k\pm} &&= \frac{kq}{2m}\pm \frac{4 \xi_k\gamma_k}{E_{k+}^2 - E_{k-}^2} \tilde{h}_z \nonumber \\ &&+\frac1{2E_{k\pm}}\left(1\pm4\frac{\gamma_k^2+h_x^2}{E_{k+}^2-E_{k-}^2}\right)\frac{\xi_kq^2}{4m}\nonumber \\
 &&+ \frac1{2E_{k\pm}}\left[1\pm 4\frac{\xi_k^2+\gamma_k^2+\Delta^2}{E_{k+}^2-E_{k-}^2} \right.  \\
 &&+ \left.
  \frac{16 \xi_k^2\gamma_k^2}{(E_{k+}^2- E_{k-}^2)^3}\left(E_{k+}^2 - E_{k-}^2\mp 4E_{k\pm}^2\right)\right]\tilde{h}_z^2 \nonumber
\end{eqnarray}
gathers the corrections to $E_{k\pm}$, up to quadratic order in $h_z$ and $q$. 

At zero temperature,  $f = - [1/(2L_{\text{nw}})] \sum' E$, where the sum runs over the positive eigenvalues of the BdG Hamiltonian~(\ref{BdG-q}) and $L_{\rm nw}$ is the nanowire length. Keeping only the terms bilinear in $\tilde{h}_z$ and $q$ in  $\delta E_{k\pm}$, we find that the supercurrent at small $h_z$ is given by Eq. (\ref{I_exact}).

In general, a finite $q$ is required for the supercurrent to be zero. The supercurrent at small $q$ and $h_z$ is given by
\begin{widetext}
\begin{multline}\label{supercurrent-suppl-general-finite-q}
 \frac{I(q)}{e/\pi} 
 = (h_z+\alpha q/2) \alpha  h_x^2 \int\limits_{-\infty}^{\infty}dk \frac{(E_{k+}+E_{k-})^2 - 4(\xi_k^2 + \Delta^2)}{E_{k+}E_{k-}(E_{k+}+E_{k-})^3} 
 +\frac{q}{8m} \int\limits_{-\infty}^{\infty}dk \left\{\frac{\xi_k\left[(E_{k+}+E_{k-})^2 - 4(\gamma_k^2+h_x^2)\right]}{E_{k+}E_{k-}(E_{k+}+E_{k-})}-2\right\}\,.
\end{multline}
\end{widetext}
Note that it is necessary to keep track of a high-energy momentum cutoff in order to derive the second term above.

 \section{Backscattering by a barrier}
\label{Sec_appendix_transmission}

In this section, we determine the normal backscattering amplitude for electrons with energies within the helical gap of a nanowire in the presence of a longitudinal magnetic field $h_x$ and a local potential barrier. This allows us to compare the numerical results of the tight-binding Hamiltonian with the prediction for a junction through helical edge states at finite backscattering, Eq.~(\ref{I_an_barrier}).

Such a nanowire is described by the Hamiltonian
\begin{equation}
 \hat{H} = \frac{\hat{p}_x^2}{2m} + V(x) - \alpha \hat{p}_x \sigma_z - h_x \sigma_x\,,
\end{equation}
where $V(x) = U\delta(x)$. We calculate the transmission $T(E)$ at energy $E$ through this barrier in the helical gap, $-|h_x| < E < |h_x|$. 

For a given $E$ in this interval, we find the following plane-wave and evanescent solutions on either side of the barrier:
\begin{eqnarray}
 \psi_+(x) &= \begin{pmatrix}
              1 \\
              \chi
             \end{pmatrix}
 e^{ipx}\,,\qquad 
 \psi_-(x) &=  \begin{pmatrix}
              \chi \\
              1
             \end{pmatrix}
e^{-ipx}
\,,\\
\phi_+(x) &= \begin{pmatrix}
              1 \\
              \eta
             \end{pmatrix}
 e^{-\kappa x}\,,\qquad 
 \phi_-(x) &= \begin{pmatrix}
              \eta \\
              1
             \end{pmatrix}
e^{\kappa x}
\end{eqnarray}
(the normalization prefactors are not important as long as they are the same for $\psi_+$ and $\psi_-$). Here, 
\begin{eqnarray}
 p =& \sqrt{2m\left( E + m\alpha^2 + E_h\right)}\,,
\\
 i\kappa =& i\sqrt{2m\left( -E - m\alpha^2 + E_h\right)}\,,
\\
 \chi =& \left({\alpha p - m\alpha^2 - E_h}\right)/{h_x}\,,
\\
 \eta =& \left({i\alpha \kappa - m\alpha^2 + E_h}\right)/{h_x} \,,
\end{eqnarray}
where
\begin{equation}
 E_h = \sqrt{h_x^2 + 2m\alpha^2 E + (m\alpha^2)^2}\,.
\end{equation}

The solutions of the entire Schr\"odinger equation to the left ($x<0$) and to the right ($x>0$) of the barrier have the following forms, respectively:
\begin{multline}
 \Psi_<(x) \\
 = A_+\begin{pmatrix}
              1 \\
              \chi
             \end{pmatrix}
 e^{ipx} + A_- \begin{pmatrix}
              \chi \\
              1
             \end{pmatrix}
e^{-ipx} + B\begin{pmatrix}
              \eta \\
              1
             \end{pmatrix}
e^{\kappa x} 
\end{multline}
and 
\begin{multline}
\Psi_>(x) \\
= C_+\begin{pmatrix}
              1 \\
              \chi
             \end{pmatrix}
 e^{ipx} + C_- \begin{pmatrix}
              \chi \\
              1
             \end{pmatrix}
e^{-ipx} + D\begin{pmatrix}
              1 \\
              \eta
             \end{pmatrix}
e^{-\kappa x}\,.
\end{multline}
Matching these solutions at $x=0$ as 
\begin{equation}
 \Psi_<(0^-) = \Psi_>(0^+) 
 \end{equation}
 and 
 \begin{multline}
 \left.\frac{d \Psi_<}{dx} \right|_{x=0^-} - \left.\frac{d \Psi_>}{dx} \right|_{x=0^+} 
 = -2mU \Psi(0) \\
 = -mU\left[\Psi_<(0^-) + \Psi_>(0^+)\right]\,,
\end{multline}
we eliminate $B$ and $D$, and find the relation $(A_-\,, C_+)^T = S (A_+ \,, C_-)^T$ between the remaining amplitudes, which defines the $2\times 2$ scattering matrix $S$. Noticing that $\eta \eta^* = 1$, we find after some algebraic manipulations:
\begin{equation}
 S = \frac{\eta}{a^2 - b^2} \begin{pmatrix}
                             |a|^2 - |b|^2 && ab^* - a^* b \\
                             ab^* -a^*b && |a|^2 - |b|^2
                            \end{pmatrix}
\,,
\end{equation}
where
\begin{equation}
 a = (ip +\kappa) (\chi\eta - 1) \quad \text{and}\quad 
 b = (-ip + \kappa +2mU) (\chi - \eta)\,.
\end{equation}
Thus, the transmission probability in the helical gap  is given by
\begin{equation}\label{eq:T}
 T(E) = \left|\frac{ab^* - a^*b}{a^2 - b^2}\right|^2\,.
\end{equation}

\end{appendix}


%

\end{document}